\documentclass[
    reprint, twocolumn,
    aps,prl,10pt,amsmath,amssymb,
    longbibliography,superscriptaddress,
    showpacs,preprintnumbers,
    groupedaddress]{revtex4-1}

\usepackage[colorlinks=true,citecolor=blue,linkcolor=blue]{hyperref}
\usepackage{oubraces}
\usepackage{bm,bbm,mathbbol}
\usepackage{physics}
\usepackage{mathtools}
\usepackage{amsmath, amsthm, amssymb, amsfonts}
\usepackage{bbold}
\usepackage{esvect}

\usepackage{epsfig}
\usepackage{array}
\usepackage{multirow}
\usepackage{graphicx}
\usepackage[normalem]{ulem}

\usepackage{float}
\usepackage[section]{placeins}
\usepackage{afterpage}

\usepackage{lipsum}

\usepackage{xifthen}

\makeatletter
\newcommand\footnoteref[1]{\protected@xdef\@thefnmark{\ref{#1}}\@footnotemark}
\makeatother

\newcommand{\nnnl}{\nonumber \\}

\newcommand{\wannier}[0]{{\sc Wannier90}}

\newcommand{\myeqref}[1]{Eq.~\eqref{#1}}

\newcommand{\mb}[1]{{\boldsymbol{\mathbf{#1}}}}

\newcommand{\bba}[0]{\mathbbm{a}}
\newcommand{\bbv}[0]{\mathbbm{v}}

\newcommand{\bbtau}[0]{\mathbbm{t}}

\begin{document}

\title{Comment on ``\textit{Ab initio} calculation of the shift photocurrent by Wannier interpolation''}

\author{Jae-Mo Lihm}
\email{jaemo.lihm@gmail.com}
\affiliation{Department of Physics and Astronomy, Seoul National University, Seoul 08826, Korea}

\date{\today}

\begin{abstract}
In a recent paper, Iba{\~n}ez-Azpiroz {\it et al.} [Phys. Rev. B \textbf{97}, 245143 (2018)] derive a band-truncation-error-free formula for calculating the generalized derivative of the interband dipole matrix using Wannier interpolation.
In practice, the denominators involving intermediate states are regularized by introducing a finite broadening parameter.
In this Comment, I show that when a finite broadening parameter is used, a correction term must be added to the generalized derivative to obtain results that are independent of the phase convention for the Bloch sums.
\end{abstract}

\maketitle

\citet{2018IbanezShift} derive a truncation-error-free formula for the Wannier interpolation of the generalized derivative of the interband dipole matrix. This result enables an efficient and accurate calculation of the shift-current response.

In practice, to avoid numerical problems due to near-degenerate states, the energy denominator involving intermediate states is regularized by introducing a broadening parameter $\eta$ (see Eq.~(38) of \citet{2018IbanezShift}).
Within the Wannier-interpolation scheme, the regularized version of the generalized derivative of the interband dipole matrix element is given from Eqs.~(36) and (38) of \citet{2018IbanezShift} by
\begin{align} \label{eq:r_wann_eta_def}
    &\widetilde{r}^{a;b}_{nm}(\eta)
    = \widetilde{r}^{a;b}_{nm}(0) + \frac{i}{\omega_{nm}} \nnnl
    &\times \sum_{p\neq n,m}^{M}
    \Bigg\{
    \Re\left[\frac{-i\eta}{\omega_{pm}(\omega_{pm} + i\eta)}\right] (\bbv_{np}^a + i \omega_{nm} \bba_{np}^a) \bbv_{pm}^b \nnnl
    & \quad- \Re\left[\frac{-i\eta}{\omega_{np}(\omega_{np} + i\eta)}\right] \bbv_{np}^b (\bbv_{pm}^a + i \omega_{nm} \bba_{pm}^a) \Bigg\},
\end{align}
where $M$ is the number of Wannier functions.
Here, $\widetilde{r}^{a;b}_{nm}(0)$ refers to the case without regularization, which equals to $r^{a;b}_{nm}$ in Eq.~(36) of \citet{2018IbanezShift}.
I use tilde to distinguish the quantity obtained from the Wannier interpolation scheme of \citet{2018IbanezShift} from a direct calculation using Kohn-Sham wavefunctions.

In this Comment, I show that when $\eta$ is nonzero, $\widetilde{r}^{a;b}_{nm}(\eta)$ defined in \myeqref{eq:r_wann_eta_def}, as well as the corresponding shift-current spectrum, depends on the phase convention for the Bloch sums.
I derive a correction term that should be added to \myeqref{eq:r_wann_eta_def} to obtain a phase-convention-independent result.

I consider two phase conventions, the tight-binding (TB) convention [Eq.~(14) of \citet{2018IbanezShift}]
\begin{equation}
    \ket{u_{\mb{k}j}^\mathrm{(W,\,TB)}} = \sum_{\mb{R}} e^{-i\mb{k}\cdot(\hat{\mb{r}} - \mb{R} - \mb{\tau}_j)} \ket{\mb{R}j}
\end{equation}
and the Wannier90 (W90) convention
\begin{equation}
    \ket{u_{\mb{k}j}^\mathrm{(W,\,W90)}} = \sum_{\mb{R}} e^{-i\mb{k}\cdot(\hat{\mb{r}} - \mb{R})} \ket{\mb{R}j}.
\end{equation}
Here, $\mb{\tau}_j$ is the position of the $j$-th atom of the unit cell.
The ``internal'' matrix elements that appear in \myeqref{eq:r_wann_eta_def} such as $\bbv$ and $\bba$ depend on the phase convention.
In the tight-binding (TB) convention, one finds
\begin{equation} \label{eq:r_dH_TB}
    \partial^a H_{ij}^{\rm (W, TB)} = \sum_\mb{R} e^{i \mb{k} \cdot (\mb{R} + \mb{\tau}_j - \mb{\tau}_i)} i (R^a + \tau^a_j - \tau^a_i ) H_{ij;\mb{R}},
\end{equation}
\begin{equation} \label{eq:r_A_TB}
    A_{ij}^{a\mathrm{(W, TB)}} = \sum_\mb{R} e^{i \mb{k} \cdot (\mb{R} + \mb{\tau}_j - \mb{\tau}_i)} r^a_{ij;\mb{R}} - \tau^a_i \delta_{ij}.
\end{equation}
In the W90 convention, one finds
\begin{equation} \label{eq:r_dH_W90}
    \partial^a H_{ij}^{\rm (W, W90)} = \sum_{\mb{R}} e^{i \mb{k} \cdot \mb{R}} i R^a H_{ij;\mb{R}},
\end{equation}
\begin{equation} \label{eq:r_A_W90}
    A_{ij}^{a\mathrm{(W, W90)}} = \sum_\mb{R} e^{i \mb{k} \cdot \mb{R}} r^a_{ij;\mb{R}}.
\end{equation}
The unitary matrices that diagonalize the Hamiltonians in the two conventions can be related as
\begin{equation}
    U^{(\mathrm{TB})}_{in} = e^{-i\mb{k}\cdot\mb{\tau}_i} U^{(\mathrm{W90})}_{in}.
\end{equation}

Comparing Eqs.~(\ref{eq:r_dH_W90}, \ref{eq:r_A_W90}) with Eqs.~(\ref{eq:r_dH_TB}, \ref{eq:r_A_TB}), one can show that the internal matrix elements in the two conventions are related as
\begin{equation} \label{eq:v_TB_vs_W90}
    \bbv^{a,(\mathrm{TB})}_{nm}
    = \bbv^{a,(\mathrm{W90})}_{nm} + i \omega_{nm} \bbtau^{a,(\mathrm{TB})}_{nm}
\end{equation}
and
\begin{equation} \label{eq:a_TB_vs_W90}
    \bba^{a,(\mathrm{TB})}_{nm}
    = \bba^{a,(\mathrm{W90})}_{nm} - \bbtau^{a,(\mathrm{TB})}_{nm}
\end{equation}
where
\begin{equation} \label{eq:r_tau_def}
    \bbtau^{a,(\mathrm{TB})}_{nm} \equiv \sum_i U^{(\mathrm{TB})\dagger}_{ni} \tau^a_i U^{(\mathrm{TB})}_{im}.
\end{equation}
Thus, if there is an atom that is not at the origin of the unit cell, the two phase conventions give different internal matrix elements.
\citet{2018IbanezShift} showed that when $\eta = 0$, the additional terms involving $\bbtau$ cancel out so that the two conventions give identical shift-current spectrum.
Here, I show that such cancellation is incomplete when $\eta \neq 0$.

First, let us define the generalized derivative with finite broadening $\eta$ based on Bloch wavefunctions as follows:
\begin{align} \label{eq:r_bloch_eta_def}
    &r^{a;b}_{nm}(\eta)
    = \frac{i}{\omega_{nm}} \Bigg\{
    \frac{v_{nm}^a \Delta_{nm}^b + v_{nm}^b \Delta_{nm}^a}{\omega_{nm}}
    - w^{ab}_{nm} \nnnl
    &+ \sum_{p\neq n,m}^{\infty} \Big[
    \Re\left(\frac{1}{\omega_{pm} + i\eta}\right) v_{np}^a v_{pm}^b \nnnl 
    &\quad - \Re\left(\frac{1}{\omega_{np} + i\eta}\right) v_{np}^b v_{pm}^a \Big]
    \Bigg\}.
\end{align}
When $\eta=0$, $r^{a;b}_{nm}(\eta)$ reduces to Eq.~(12) of \citet{2018IbanezShift}.
Our goal is to obtain a Wannier interpolation formula for $r^{a;b}_{nm}(\eta)$ which holds true at finite $\eta$.

We find
\begin{align} \label{eq:r_bloch_diff}
    r^{a;b}_{nm}(\eta) - r^{a;b}_{nm}(0)
    = \frac{i}{\omega_{nm}} \sum_{p\neq n,m}^{\infty} A_{nm,p}^{a,b}(\eta)
\end{align}
where we defined
\begin{align} \label{eq:r_diff_mel}
    A_{nm,p}^{a,b}(\eta)
    &= \Re\left(\frac{1}{\omega_{pm} + i\eta} - \frac{1}{\omega_{pm}}\right) v_{np}^a v_{pm}^b \nnnl
    &- \Re\left(\frac{1}{\omega_{np} + i\eta} - \frac{1}{\omega_{np}}\right) v_{np}^b v_{pm}^a.
\end{align}
In principle, \myeqref{eq:r_bloch_diff} requires a sum over an infinite number of unoccupied bands and thus cannot be computed using Wannier interpolation.
However, note that
\begin{align}
    \Re\left(\frac{1}{\omega_{pm} + i\eta} - \frac{1}{\omega_{pm}}\right)
    &= \Re\left[\frac{-i\eta}{\omega_{pm}(\omega_{pm} + i\eta)}\right] \nnnl
    &= -\frac{\eta^2}{\omega_{pm}(\omega_{pm}^2 + \eta^2)}
\end{align}
is very small as long as $\abs{\omega_{pm}}$ is much larger than $\eta$.
Also, when computing the shift-current response, $r^{a;b}_{nm}$ is computed only for the low-energy states $n$ and $m$ which are a pair of occupied and unoccupied states that can be resonantly excited by an external electric field.
Thus, \myeqref{eq:r_bloch_diff} converges rapidly with the number of states included in the sum.
We exploit this rapid convergence to approximate the infinite sum in \myeqref{eq:r_bloch_diff} by a finite sum over Wannier-interpolated bands
\begin{equation} \label{eq:r_bloch_diff_wann}
    r^{a;b}_{nm}(\eta) - r^{a;b}_{nm}(0)
    \approx \frac{i}{\omega_{nm}} \sum_{p\neq n,m}^{M} A_{nm,p}^{a,b}(\eta),
\end{equation}
where $A_{nm,p}^{a,b}(\eta)$ [\myeqref{eq:r_diff_mel}] is calculated using the Wannier interpolation of energy and velocity matrix elements.
Later, I demonstrate the validity of this approximation using numerical calculation (see Fig.~\ref{fig:truncation}).

Now, let us study the relation between \myeqref{eq:r_wann_eta_def} and \myeqref{eq:r_bloch_eta_def}.
The velocity matrix element can be written as
\begin{equation}
    v_{nm}^a = \bbv_{nm}^a + i \omega_{nm} \bba_{nm}^a.
\end{equation}
This expression can be obtained from Eqs.~(11), (22), and (23) of \citet{2018IbanezShift}.
From Eqs.~(\ref{eq:v_TB_vs_W90}, \ref{eq:a_TB_vs_W90}), one can easily show that $v_{nm}^a$ does not depend on the phase convention.
By using Eqs.~(\ref{eq:r_wann_eta_def}, \ref{eq:r_diff_mel}), we find
\begin{widetext}
\begin{align} \label{eq:A_using_tilde}
    \frac{i}{\omega_{nm}} \sum_{p\neq n,m}^{M} A_{nm,p}^{a,b}(\eta)
    =& \widetilde{r}^{a;b}_{nm}(\eta) - \widetilde{r}^{a;b}_{nm}(0)
    + \frac{i}{\omega_{nm}} \sum_{p\neq n,m}^{M}
    \Re\left[\frac{-i\eta}{\omega_{pm}(\omega_{pm} + i\eta)}\right]
    \left[ v^a_{np} v^b_{pm} - (\bbv_{np}^a + i \omega_{nm} \bba_{np}^a) \bbv_{pm}^b \right] \nnnl
    &\hskip 7.2em - \frac{i}{\omega_{nm}} \sum_{p\neq n,m}^{M}
    \Re\left[\frac{-i\eta}{\omega_{np}(\omega_{np} + i\eta)}\right]
    \left[  v^b_{np} v^a_{pm} - \bbv_{np}^b (\bbv_{pm}^a + i \omega_{nm} \bba_{pm}^a) \right] \nnnl
    =& \widetilde{r}^{a;b}_{nm}(\eta) - \widetilde{r}^{a;b}_{nm}(0)
    + \frac{i}{\omega_{nm}} \sum_{p\neq n,m}^{M}
    \Re\left[\frac{-i\eta}{\omega_{pm}(\omega_{pm} + i\eta)}\right]
    \left[ i \omega_{mp} \bba_{np}^a \bbv_{pm}^b + v^a_{np} i \omega_{pm} \bba_{pm}^b \right] \nnnl
    &\hskip 7.2em - \frac{i}{\omega_{nm}} \sum_{p\neq n,m}^{M}
    \Re\left[\frac{-i\eta}{\omega_{np}(\omega_{np} + i\eta)}\right]
    \left[ i \omega_{pn} \bbv_{np}^b \bba_{pm}^a + i \omega_{np} \bba_{np}^b v^a_{pm} \right] \nnnl
    =& \widetilde{r}^{a;b}_{nm}(\eta) - \widetilde{r}^{a;b}_{nm}(0)
    + \frac{1}{\omega_{nm}} \sum_{p\neq n,m}^{M}
    \Re\left(\frac{-i\eta}{\omega_{pm} + i\eta}\right) \left[ \bba_{np}^a \bbv_{pm}^b
    - (\bbv_{np}^a + i \omega_{np} \bba_{np}^a)
    \bba_{pm}^b \right] \nnnl
    &\hskip 7.2em - \frac{1}{\omega_{nm}} \sum_{p\neq n,m}^{M}
    \Re\left(\frac{-i\eta}{\omega_{np} + i\eta}\right) \left[
    \bbv_{np}^b \bba_{pm}^a
    - \bba_{np}^b
    (\bbv_{pm}^a + i \omega_{pm} \bba_{pm}^a) \right].
\end{align}

Using Eqs.~(\ref{eq:r_bloch_diff_wann}, \ref{eq:A_using_tilde}) and $r^{a;b}_{nm}(0) = \widetilde{r}^{a;b}_{nm}(0)$ which is proven in \citet{2018IbanezShift}, we find
\begin{align} \label{eq:genr_diff_corr}
    r^{a;b}_{nm}(\eta)
    \approx \widetilde{r}^{a;b}_{nm}(\eta)
    -& \frac{1}{\omega_{nm}} \sum_{p\neq n,m}^{M}
    \frac{\eta^2}{\omega_{pm}^2 + \eta^2} \left[ \bba_{np}^a \bbv_{pm}^b
    - (\bbv_{np}^a + i \omega_{np} \bba_{np}^a)
    \bba_{pm}^b \right] \nnnl
    +& \frac{1}{\omega_{nm}} \sum_{p\neq n,m}^{M}
    \frac{\eta^2}{\omega_{np}^2 + \eta^2} \left[
    \bbv_{np}^b \bba_{pm}^a
    - \bba_{np}^b
    (\bbv_{pm}^a + i \omega_{pm} \bba_{pm}^a) \right].
\end{align}
\end{widetext}
The right-hand side of \myeqref{eq:genr_diff_corr} is the desired phase-convention-independent Wannier interpolation formula for the generalized derivative of the interband dipole matrix element at finite $\eta$.
This expression is independent of the phase convention.
The reason is as follows.
Equation~\eqref{eq:A_using_tilde} is phase-convention independent because $A^{a,b}_{nm,p}(\eta)$ [\myeqref{eq:r_diff_mel}] is defined only in terms of phase-convention independent matrix elements.
Also, $\widetilde{r}^{a;b}_{nm}(0)$ is phase-convention independent because it equals $r^{a;b}_{nm}(0)$ which is defined in terms of Kohn-Sham wavefunctions.
Since the right-hand side of \myeqref{eq:genr_diff_corr} is the sum of \myeqref{eq:A_using_tilde} and $\widetilde{r}^{a;b}_{nm}(0)$, it is also independent of the phase convention.

When $\eta = 0$, the correction terms in \myeqref{eq:genr_diff_corr} vanish as expected.
When $\eta$ is nonzero, the correction terms are finite.
Also, we find that these correction terms depend on the phase convention used for the Bloch sums.
Thus, if this correction is not applied, the calculated $r^{a;b}_{nm}(\eta)$ matrix elements, as well as the shift-current spectrum, becomes dependent on the phase convention.

\begin{figure}
\includegraphics[width=1.0\columnwidth]{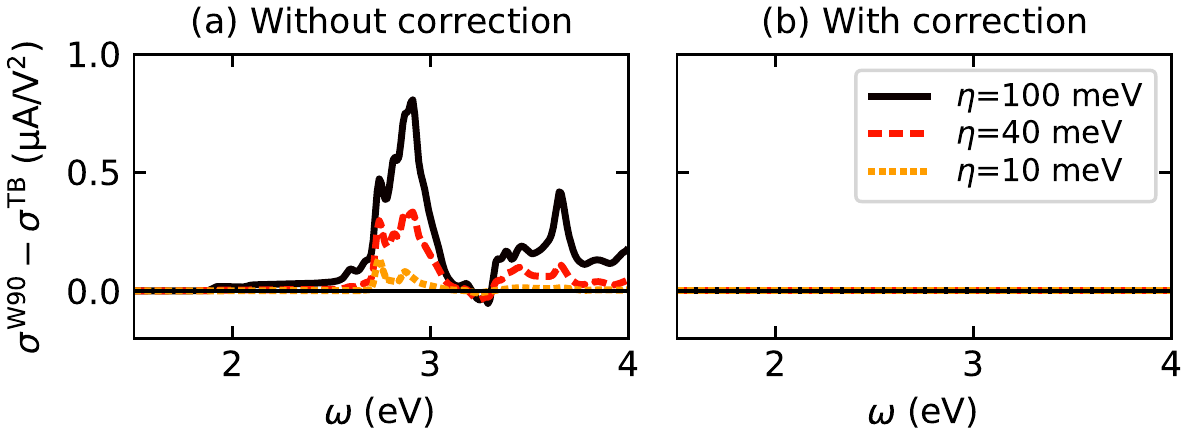}
\caption{
Difference of the $zzz$ component of the shift-current spectrum of monolayer GeS computed using the W90 and TB conventions (a) without and (b) with the finite-$\eta$ correction [\myeqref{eq:genr_diff_corr}].
}
\label{fig:sc_diff}
\end{figure}

\begin{figure}
\includegraphics[width=1.0\columnwidth]{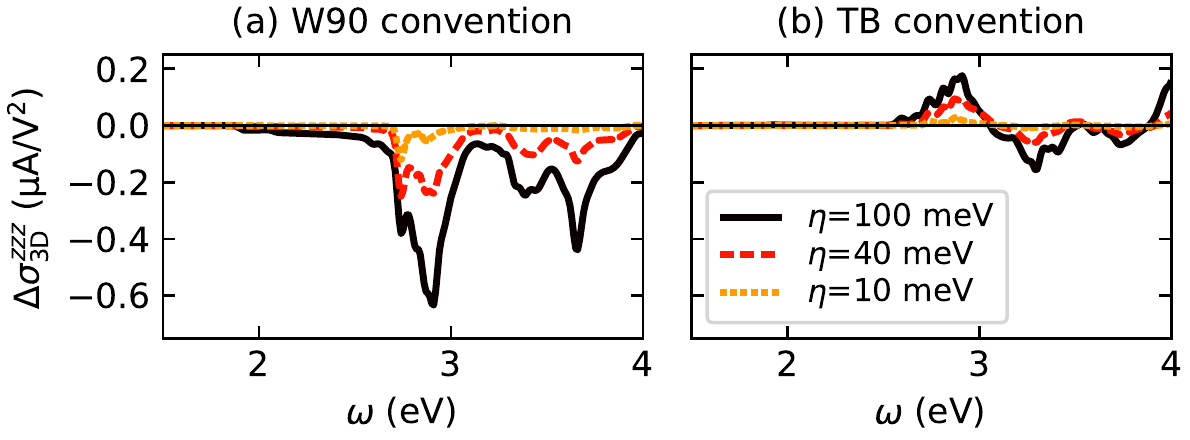}
\caption{
Change in the shift-current spectrum of monolayer GeS computed with the (a) W90 convention and the (b) TB convention due to the finite-$\eta$ correction: $\Delta \sigma = \sigma^{\rm w/ corr.} - \sigma^{\rm w/o\ corr.}$.
}
\label{fig:sc_corr}
\end{figure}

Now, I demonstrate these findings with numerical calculations.
I implemented the correction formula \myeqref{eq:genr_diff_corr} in \wannier\ and calculated the shift-current spectrum of monolayer GeS.
For the density functional theory calculation, I used fully relativistic ONCV pseudopotentials~\cite{2013HamannONCVPSP} taken from the PseudoDojo library (v0.4)~\cite{2018VanSettenPseudoDojo} and a wavefunction kinetic energy cutoff of 80~Ry.
Other computational parameters are the same as in \citet{2018IbanezShift} unless otherwise noted.

Figure~\ref{fig:sc_diff} shows the difference between the shift-current spectra calculated using the two conventions.
Without the correction term, there is a small but nonzero discrepancy between the two spectra.
The discrepancy increases with $\eta$.
When the correction term is added, the shift-current spectra calculated using the TB and the W90 conventions perfectly agree for all values of $\eta$.
This result demonstrates the phase-convention independence of \myeqref{eq:genr_diff_corr}.

Figure~\ref{fig:sc_corr} shows the change in the shift-current spectrum due to the correction term.
For $\eta$=40~meV, which is used in \citet{2018IbanezShift}, the change in the shift-current spectrum due to the correction is below 0.3~$\mathrm{\mu A/V^2}$.
This correction is nonzero but will have little effect on the calculated shift-current spectrum [Fig.~5(a) of \citet{2018IbanezShift}] whose magnitude is around 50~$\mathrm{\mu A/V^2}$.

Finally, I test the validity of approximating the infinite sum over bands in \myeqref{eq:r_bloch_diff_wann} by a finite sum over Wannier-interpolated bands as \myeqref{eq:r_bloch_diff_wann}.
Let us define the following approximant of \myeqref{eq:r_bloch_diff}:
\begin{equation} \label{eq:r_bloch_diff_approx}
    f_{n,m}^{a;b}(E_{\rm cutoff};\eta)
    = \frac{i}{\omega_{nm}} \sum_{p\neq n,m}^{\varepsilon_p \le E_{\rm cutoff}}
    A_{nm,p}^{a,b}(\eta).
\end{equation}
Here, I replaced the infinite sum over $p$ in \myeqref{eq:r_bloch_diff} with a finite sum for states with $\varepsilon_p \le E_{\rm cutoff}$.
Equation~\eqref{eq:r_bloch_diff} is recovered in the limit $E_{\rm cutoff} \rightarrow \infty$.
In a Wannier interpolation, only the bands inside the frozen window for Wannierization are accurately interpolated.
So, to use Wannier interpolation, we choose $E_{\rm cutoff}$ to be $E_{\rm froz.}$, the upper bound of the frozen window for Wannierization.
Then, the truncation error in evaluating \myeqref{eq:r_bloch_diff} is
\begin{equation} \label{eq:r_bloch_trunc}
    \delta r^{a;b}_{nm}(\eta)
    = f_{n,m}^{a;b}(\infty;\eta) - f_{n,m}^{a;b}(E_{\rm froz.};\eta).
\end{equation}

\begin{figure}
\includegraphics[width=1.0\columnwidth]{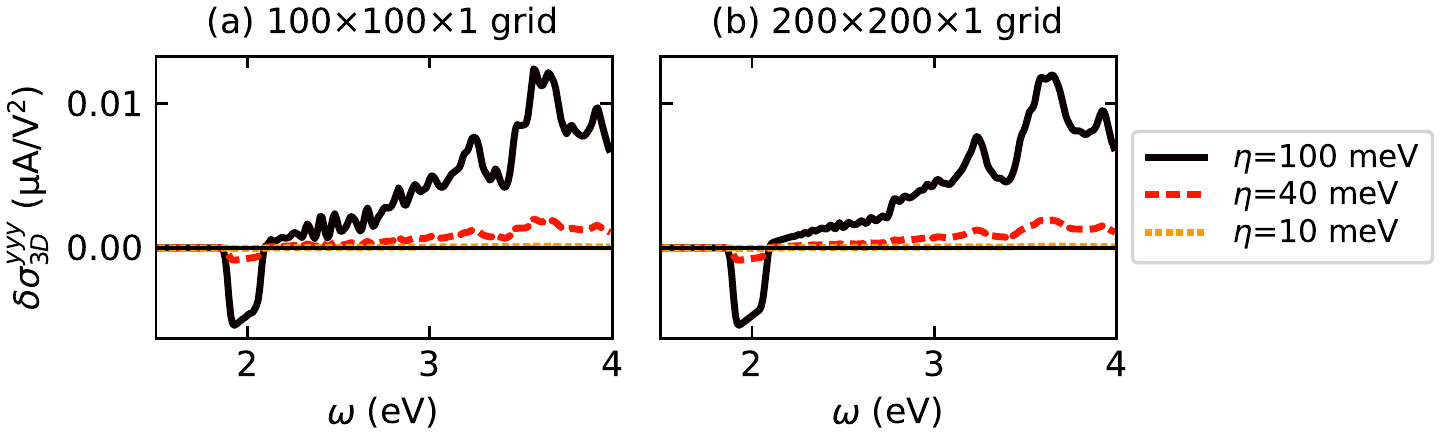}
\caption{
Truncation error in the finite-$\eta$ correction of the shift current spectrum of monolayer GeS calculated by replacing $r^{a;b}_{nm}$ in the shift current formula [Eqs.~(5, 8) of \cite{2018IbanezShift}] with $\delta r^{a;b}_{nm}(\eta)$ [\myeqref{eq:r_bloch_trunc}].
We used a (a) 100$\times$100$\times$1 and a (b) 200$\times$200$\times$1 $k$-point grid.
}
\label{fig:truncation}
\end{figure}

To study the magnitude of the error in the shift-current spectra due to this truncation, I calculated the shift-current spectrum with $r^{a;b}_{nm}$ in the shift current formula [Eqs.~(5, 8) of \cite{2018IbanezShift}] replaced with $\delta r^{a;b}_{nm}(\eta)$ [\myeqref{eq:r_bloch_trunc}].
Since the high-energy bands need to be explicitly included, I performed the calculation at the DFT level.
The infinity in \myeqref{eq:r_bloch_trunc} was replaced with the conduction band minimum energy plus 20~eV, as the contribution from bands with even higher energies was found to be negligible.
Due to the high computational cost, I used $k$-point grids coarser than the 1000$\times$1000$\times$1 grid used in the calculation of the shift-current spectra.
Figure~\ref{fig:truncation} shows the truncation error of the shift-current spectra.
The results indicate that the magnitude of the truncation error is almost converged at a 200$\times$200$\times$1 grid.
The truncation error of the shift-current spectra is at most 0.012~$\mathrm{\mu A/V^2}$, even at a large $\eta$ of 100~meV up to $\omega$=4~eV when the upper limit of the frozen window was set to be 5~eV above the top of the valence band~\cite{2018IbanezShift}.
For each $\eta$, the truncation error is much smaller than the finite-$\eta$ correction of the shift-current spectrum shown in Fig.~\ref{fig:sc_corr}.
This analysis shows that the states with energy above the frozen window give negligible contributions to the finite-$\eta$ correction of the shift-current spectra.
Therefore, we can safely approximate the infinite sum in \myeqref{eq:r_bloch_diff} by a sum over Wannier-interpolated bands as in \myeqref{eq:r_bloch_diff_wann}.

Note that the use of $\eta$ is necessary to avoid numerical problems due to near-degenerate states.
As one uses smaller $\eta$, a larger number of $k$ points need to be used to converge the calculated shift-current spectrum.
In other words, the number of $k$ points used gives the lower bound of $\eta$ for the calculated spectrum to remain stable.

I also note that I fixed a small inconsistency between the shift current implementation of \wannier\ and the formalism of \citet{2018IbanezShift}.
In the \wannier\ implementation, the broadening $\eta$ was included in denominators that do not contain any intermediate states, such as $1/\omega_{nm}$ in \myeqref{eq:r_wann_eta_def}, while the formalism~\cite{2018IbanezShift} does not include $\eta$ in such cases.
I changed the \wannier\ code so that it is consistent with \citet{2018IbanezShift}.
Without this change, the exact agreement between the two phase conventions was not reached.
All results in this Comment were obtained after this change.

To summarize, I showed that when using a finite broadening parameter $\eta$ to calculate the generalized derivative of the interband dipole matrix, one needs to add a correction term [\myeqref{eq:genr_diff_corr}] to Eq.~(36) of \citet{2018IbanezShift}.
Otherwise, the calculated shift-current spectrum becomes dependent on the phase convention for the Bloch sums.

\begin{acknowledgments}
Computational resources were provided by KISTI Supercomputing Center (KSC-2020-INO-0078).
\end{acknowledgments}

\bibliography{main}

\end{document}